\documentclass{PoS}
\usepackage{ifthen} 
\newboolean{pdflatex}
\setboolean{pdflatex}{true} 

\newboolean{articletitles}
\setboolean{articletitles}{true} 

\newboolean{uprightparticles}
\setboolean{uprightparticles}{false} 

\newboolean{inbibliography}
\setboolean{inbibliography}{false} 

\usepackage{microtype}
\usepackage{lineno}  
\usepackage{xspace} 
\usepackage{caption} 

\usepackage{graphicx}  
\usepackage{color}
\usepackage{subcaption}
\usepackage{colortbl}
\graphicspath{{./figs/}} 

\usepackage{amsmath} 
\usepackage{amssymb}
\usepackage{amsfonts}
\usepackage{upgreek} 

\newcommand*\patchAmsMathEnvironmentForLineno[1]{%
\expandafter\let\csname old#1\expandafter\endcsname\csname #1\endcsname
\expandafter\let\csname oldend#1\expandafter\endcsname\csname
end#1\endcsname
 \renewenvironment{#1}%
   {\linenomath\csname old#1\endcsname}%
   {\csname oldend#1\endcsname\endlinenomath}%
}
\newcommand*\patchBothAmsMathEnvironmentsForLineno[1]{%
  \patchAmsMathEnvironmentForLineno{#1}%
  \patchAmsMathEnvironmentForLineno{#1*}%
}
\AtBeginDocument{%
\patchBothAmsMathEnvironmentsForLineno{equation}%
\patchBothAmsMathEnvironmentsForLineno{align}%
\patchBothAmsMathEnvironmentsForLineno{flalign}%
\patchBothAmsMathEnvironmentsForLineno{alignat}%
\patchBothAmsMathEnvironmentsForLineno{gather}%
\patchBothAmsMathEnvironmentsForLineno{multline}%
\patchBothAmsMathEnvironmentsForLineno{eqnarray}%
}

\usepackage{xspace} 
\usepackage{upgreek}

\def\MagUp {\mbox{\em Mag\kern -0.05em Up}\xspace}

\ifthenelse{\boolean{uprightparticles}}%
{

 \def\Pnu         {\ensuremath{\upnu}\xspace}                 
                  
 \def\Ppi         {\ensuremath{\uppi}\xspace}

 \def\PDelta      {\ensuremath{\Delta}\xspace}                 
 \def\PXi      {\ensuremath{\Xi}\xspace}                 
 \def\PLambda      {\ensuremath{\Lambda}\xspace}                 
 \def\PSigma      {\ensuremath{\Sigma}\xspace}                 
 \def\POmega      {\ensuremath{\Omega}\xspace}                 
 \def\PUpsilon      {\ensuremath{\Upsilon}\xspace}                 
 
 \def\PB      {\ensuremath{\mathrm{B}}\xspace}                 
                  
 \def\PD      {\ensuremath{\mathrm{D}}\xspace}

 \def\PK      {\ensuremath{\mathrm{K}}\xspace}

 \def\Pi      {\ensuremath{\mathrm{i}}\xspace}

}
{

 \def\Pnu         {\ensuremath{\nu}\xspace}                 
                  
 \def\Ppi         {\ensuremath{\pi}\xspace}

 \mathchardef\PDelta="7101
 \mathchardef\PXi="7104
 \mathchardef\PLambda="7103
 \mathchardef\PSigma="7106
 \mathchardef\POmega="710A
 \mathchardef\PUpsilon="7107
                  
 \def\PB      {\ensuremath{B}\xspace}                 
                  
 \def\PD      {\ensuremath{D}\xspace}

 \def\PK      {\ensuremath{K}\xspace}

 \def\Pi      {\ensuremath{i}\xspace}

}

\makeatletter
\ifcase \@ptsize \relax
  \newcommand{\miniscule}{\@setfontsize\miniscule{4}{5}}
\or
  \newcommand{\miniscule}{\@setfontsize\miniscule{5}{6}}
\or
  \newcommand{\miniscule}{\@setfontsize\miniscule{5}{6}}
\fi
\makeatother

\DeclareRobustCommand{\optbar}[1]{\shortstack{{\miniscule (\rule[.5ex]{1.25em}{.18mm})}
  \\ [-.7ex] $#1$}}

\def\neub       {{\ensuremath{\overline{\Pnu}}}\xspace}

\def\neumb      {{\ensuremath{\neub_\mu}}\xspace}

\def\pion   {{\ensuremath{\Ppi}}\xspace}

\def\pip    {{\ensuremath{\pion^+}}\xspace}
\def\pim    {{\ensuremath{\pion^-}}\xspace}

\def\kaon    {{\ensuremath{\PK}}\xspace}
  \def\Kbar    {{\kern 0.2em\overline{\kern -0.2em \PK}{}}\xspace}

\def\KorKbar    {\kern 0.18em\optbar{\kern -0.18em K}{}\xspace}

\def\Kzb     {{\ensuremath{\Kbar{}^0}}\xspace}
\def\Kp      {{\ensuremath{\kaon^+}}\xspace}
\def\Km      {{\ensuremath{\kaon^-}}\xspace}

  \def\Dbar    {{\kern 0.2em\overline{\kern -0.2em \PD}{}}\xspace}
\def\D       {{\ensuremath{\PD}}\xspace}

\def\DorDbar    {\kern 0.18em\optbar{\kern -0.18em D}{}\xspace}
\def\Dz      {{\ensuremath{\D^0}}\xspace}
\def\Dzb     {{\ensuremath{\Dbar{}^0}}\xspace}
\def\Dp      {{\ensuremath{\D^+}}\xspace}

\def\Dstarp  {{\ensuremath{\D^{*+}}}\xspace}

\def\Bbar    {{\ensuremath{\kern 0.18em\overline{\kern -0.18em \PB}{}}}\xspace}
\def\Bb      {{\ensuremath{\Bbar}}\xspace}
\def\BorBbar    {\kern 0.18em\optbar{\kern -0.18em B}{}\xspace}

  \def\Y#1S{\ensuremath{\PUpsilon{(#1S)}}\xspace}

\def\Lbar        {{\ensuremath{\kern 0.1em\overline{\kern -0.1em\PLambda}}}\xspace}
\def\LorLbar    {\kern 0.18em\optbar{\kern -0.18em \PLambda}{}\xspace}

\def\ra                 {\ensuremath{\rightarrow}\xspace}
\def\to                 {\ensuremath{\rightarrow}\xspace}

\def\CP                {{\ensuremath{C\!P}}\xspace}

\def\AT#1     {\ensuremath{A_{\mathrm{T}}^{#1}}\xspace}           

\def\C#1      {\ensuremath{\mathcal{C}_{#1}}\xspace}                       
\def\Cp#1     {\ensuremath{\mathcal{C}_{#1}^{'}}\xspace}                    
\def\Ceff#1   {\ensuremath{\mathcal{C}_{#1}^{\mathrm{(eff)}}}\xspace}        
\def\Cpeff#1  {\ensuremath{\mathcal{C}_{#1}^{'\mathrm{(eff)}}}\xspace}       
\def\Ope#1    {\ensuremath{\mathcal{O}_{#1}}\xspace}                       
\def\Opep#1   {\ensuremath{\mathcal{O}_{#1}^{'}}\xspace}                    

\newcommand{\tev}{\ifthenelse{\boolean{inbibliography}}{\ensuremath{~T\kern -0.05em eV}\xspace}{\ensuremath{\mathrm{\,Te\kern -0.1em V}}}\xspace}
\newcommand{\gev}{\ensuremath{\mathrm{\,Ge\kern -0.1em V}}\xspace}
\newcommand{\mev}{\ensuremath{\mathrm{\,Me\kern -0.1em V}}\xspace}
\newcommand{\kev}{\ensuremath{\mathrm{\,ke\kern -0.1em V}}\xspace}
\newcommand{\ev}{\ensuremath{\mathrm{\,e\kern -0.1em V}}\xspace}
\newcommand{\gevc}{\ensuremath{{\mathrm{\,Ge\kern -0.1em V\!/}c}}\xspace}
\newcommand{\mevc}{\ensuremath{{\mathrm{\,Me\kern -0.1em V\!/}c}}\xspace}
\newcommand{\gevcc}{\ensuremath{{\mathrm{\,Ge\kern -0.1em V\!/}c^2}}\xspace}
\newcommand{\gevgevcccc}{\ensuremath{{\mathrm{\,Ge\kern -0.1em V^2\!/}c^4}}\xspace}
\newcommand{\mevcc}{\ensuremath{{\mathrm{\,Me\kern -0.1em V\!/}c^2}}\xspace}

\def\invfb   {\ensuremath{\mbox{\,fb}^{-1}}\xspace}

\newcommand{\stat}{\ensuremath{\mathrm{\,(stat)}}\xspace}
\newcommand{\syst}{\ensuremath{\mathrm{\,(syst)}}\xspace}

\def\gsim{{~\raise.15em\hbox{$>$}\kern-.85em
          \lower.35em\hbox{$\sim$}~}\xspace}
\def\lsim{{~\raise.15em\hbox{$<$}\kern-.85em
          \lower.35em\hbox{$\sim$}~}\xspace}

\def\tell1  {TELL1\xspace}
\def\ukl1   {UKL1\xspace}

\usepackage{cite} 
\usepackage{mciteplus}

\usepackage{longtable} 

\newcommand{\DACP}{\ensuremath{\Delta A_{\CP}}\xspace}
\newcommand{\KK}{\ensuremath{\Dz\to\Km\Kp}\xspace}
\newcommand{\PiPi}{\ensuremath{\Dz\to\pim\pip}\xspace}
\newcommand{\KPi}{\ensuremath{\Dz\to\Km\pip}\xspace}
\newcommand{\KPiPi}{\ensuremath{\Dp\to\Km\pip\pip}\xspace}

\newcommand{\KzbPi}{\ensuremath{\Dp\to\Kzb\pip}\xspace}
\newcommand{\deltam}{{\ensuremath{\delta m}}\xspace}

\newcommand{\acpkk}{\ensuremath{A_{\CP}(\Km\Kp)}\xspace}
\newcommand{\acppipi}{\ensuremath{A_{\CP}(\pim\pip)}\xspace}

\newcommand{\AcpPiPi}{0.24}
\newcommand{\AcpPiPiStatErr}{0.15}
\newcommand{\AcpPiPiSysErr}{0.11}
\newcommand{\AcpLhcbPiPi}{0.07}
\newcommand{\AcpLhcbPiPiStatErr}{0.14}
\newcommand{\AcpLhcbPiPiSysErr}{0.11}
\newcommand{\AcpLHCb}{0.04}
\newcommand{\AcpLHCbStatErr}{0.12}
\newcommand{\AcpLHCbSysErr}{0.10}

\title{Measurement of $C\!P$ asymmetries in $D^0\rightarrow hh$ decays}

\ShortTitle{\CP asymmetries in $\Dz\to hh$ decays}

\author{\speaker{Evelina Gersabeck }\thanks{on behalf of the LHCb collaboration}\\
        Ruprecht-Karls-Universit\"at Heidelberg, Heidelberg, Germany\\
        E-mail: \email{Evelina.Gersabeck@cern.ch}}

\abstract{
          The latest measurements of the individual time-integrated \CP asymmetry in the Cabibbo-suppressed decays $\Dz\to hh$ decays are presented. The results are based on $pp$ collision data, corresponding to an integrated luminosity of 3\invfb, collected with the LHCb detector at centre-of-mass energies of 7 and 8\tev. 
    A combination of the asymmetries using both prompt and secondary charm decays is presented      
\begin{align*}
\acpkk=(\AcpLHCb\pm\AcpLHCbStatErr\stat \pm\AcpLHCbSysErr\syst)\%, \\
\acppipi=(\AcpLhcbPiPi\pm\AcpLhcbPiPiStatErr\stat \pm\AcpLhcbPiPiSysErr\syst)\%.
\end{align*}

These are the most precise measurements from a single experiment. The result for \acpkk is the most precise determination of a time-integrated \CP asymmetry in the charm sector to date, and neither measurement shows evidence of \CP asymmetry.
}
\FullConference{VIII International Workshop On Charm Physics\\
		5-9 September, 2016\\
		Bologna, Italy}

\begin{document}

Charge-parity (\CP) violation in weak decays is described by the Standard Model (SM) and is governed by a single complex phase in the Cabibbo-Kobayashi-Maskawa (CKM) matrix. The charmed $D$ meson system is suitable for probing the \CP violation in the up-type quark sector.
Recent studies of \CP violation in weak decays of $D$ mesons have not found evidence of \CP violation, while the phenomenon is well established in decays of $K$- and $B$-meson systems~\cite{Christenson:1964fg,Aubert:2001nu,Abe:2001xe,LHCb-PAPER-2013-018,LHCb-PAPER-2012-001}. 

To distinguish the two $\CP$-conjugate decays, the  flavour of the $\Dz$ at production must be known.  
At LHCb, two different techniques are employed. The  flavour of the $\Dz$ can be tagged by the charge of the soft pion, $\pi^+_{s}$, in the strong decay $\Dstarp\to\Dz\pi^+_{s}$ for prompt $\Dz$ decays~\cite{LHCb-PAPER-2015-055}, or by the charge of the muon in the semileptonic decay $\Bb\to\Dz \mu^- \neumb X$ for secondary charm decays~\cite{LHCb-PAPER-2014-013}.
The time-integrated \CP asymmetry in the decay rates of the singly Cabibbo-supressed process $\Dz \rightarrow K^-K^+$ 
\begin{equation}
A_{\CP}(\Dz\to K^-K^+)\equiv \frac{\Gamma(D^0\rightarrow K^-K^+)-\Gamma(\Dzb\rightarrow K^-K^+)}{\Gamma(D^0\rightarrow K^-K^+)+\Gamma(\Dzb\rightarrow K^-K^+)},
\end{equation}
was measured recently at LHCb~\cite{LHCb-PAPER-2016-035-mine}. This observable cannot be accessed directly due to the presence of production asymmetry of the \Dz and \Dzb mesons at $pp$ collider, and due to the presence of detection asymmetries. The latter can origin from an asymmetric interaction of the charged particles with the material of the detector. One example for this is the different cross-section for interaction with matter of \Kp and \Km. Any asymmetry in the detector e.g. some inefficient regions, faulty electronics, etc could also contribute to the detection asymmetry. The effect of such asymmetries can cancel to first order with the regular swap of the magnetic field of the dipole magnet at LHCb.

The measured raw asymmetry in the number of background-subtracted number of signal decays 
\begin{align}
A_{\rm raw} \equiv \frac{N(\Dz\to\Km\Kp)-N(\Dzb\to\Km\Kp)}{N(\Dz\to\Km\Kp)+N(\Dzb\to\Km\Kp)},
\end{align}
is related to the observable of interest, \acpkk,  via
\begin{align}
A_{\CP}(D^0\rightarrow K^-K^+) \approx A_{\rm raw}(D^0\rightarrow K^-K^+)-A_P(D^{*+})-A_D(\pi^+_{s}).
\label{eq:AsymsKK}
\end{align}
This approximation is valid only for small asymmetries, this is why a particular care has been taken to exclude regions with large asymmetries, related to detector acceptance effects~\cite{LHCb-PAPER-2015-055, LHCb-PAPER-2016-035-mine}.

To cancel the effect of the nuisance detection and production asymmetry, a set of Cabibbo-favoured decays,  $D^0\rightarrow K^-\pi^+, D^+\rightarrow K^-\pi^+\pi^+, D^+\rightarrow \Kzb\pi^+$, in which no \CP violation is expected, is used. For example, by exploiting the prompt $D^0\rightarrow K^-\pi^+$ channel, one can cancel the $\Dstarp$ production asymmetry, and the soft pion detection asymmetry. However, a new detection asymmetry is present due to the asymmetric final state $K^-\pi^+$. The effect of this asymmetry is mitigated using the other two calibration channels leading to
\begin{eqnarray}
\label{eq:AsymComb}
A_{\CP}(D^0\rightarrow K^-K^+)&=&A_{\rm raw}(D^0\rightarrow K^-K^+)-A_{\rm raw}(D^0\rightarrow K^-\pi^+) \\ 
&+&A_{\rm raw}(D^+\rightarrow K^-\pi^+\pi^+)-A_{\rm raw}(D^+\rightarrow \Kzb\pi^+) \nonumber \\
&+&A_D(\Kzb). \nonumber
\end{eqnarray}
\acpkk now only depends on measurable raw asymmetries and the calculable $\Kzb$ detection asymmetry~\cite{LHCb-PAPER-2013-003}.
The method to determine \acpkk has been first used in Ref.~\cite{LHCb-PAPER-2014-013}.

The results are obtained using a data sample of proton-proton ($pp$) collisions at centre-of-mass energies of 7 and 8$\mathrm{\,Te\kern -0.1em V}$ collected by the LHCb detector in 2011 and 2012, corresponding to approximately 3${\mbox{\,fb}^{-1}}$ of integrated luminosity.

The LHCb 
detector~\cite{Alves:2008zz,LHCb-DP-2014-002} is a
single-arm forward spectrometer covering the pseudorapidity range $2 < \eta < 5$, designed for
the study of particles containing b or c quarks. The detector elements that are particularly
relevant to this analysis are: a silicon-strip vertex detector surrounding the pp interaction
region that allows c- and b-hadrons to be identified from their characteristically long
flight distance; a tracking system that provides a measurement of momentum, $p$, of charged
particles; and two ring-imaging Cherenkov detectors that are able to discriminate between
different species of charged hadrons.

The production and detection asymmetries depend on the kinematics of the particles involved.
If the kinematic distributions are very different, this may lead to 
an imperfect cancellation of the nuisance asymmetries in \acpkk. This effect is alleviated by equalising the corresponding kinematical distributions by assigning weights to each candidate~\cite{CDF:2012qw}.

\begin{figure}[pb]

\vspace{0em}

\centering
\begin{subfigure}{0.49\textwidth}
\includegraphics[scale=0.35,  trim = 0mm 0mm 0mm 0mm, clip]{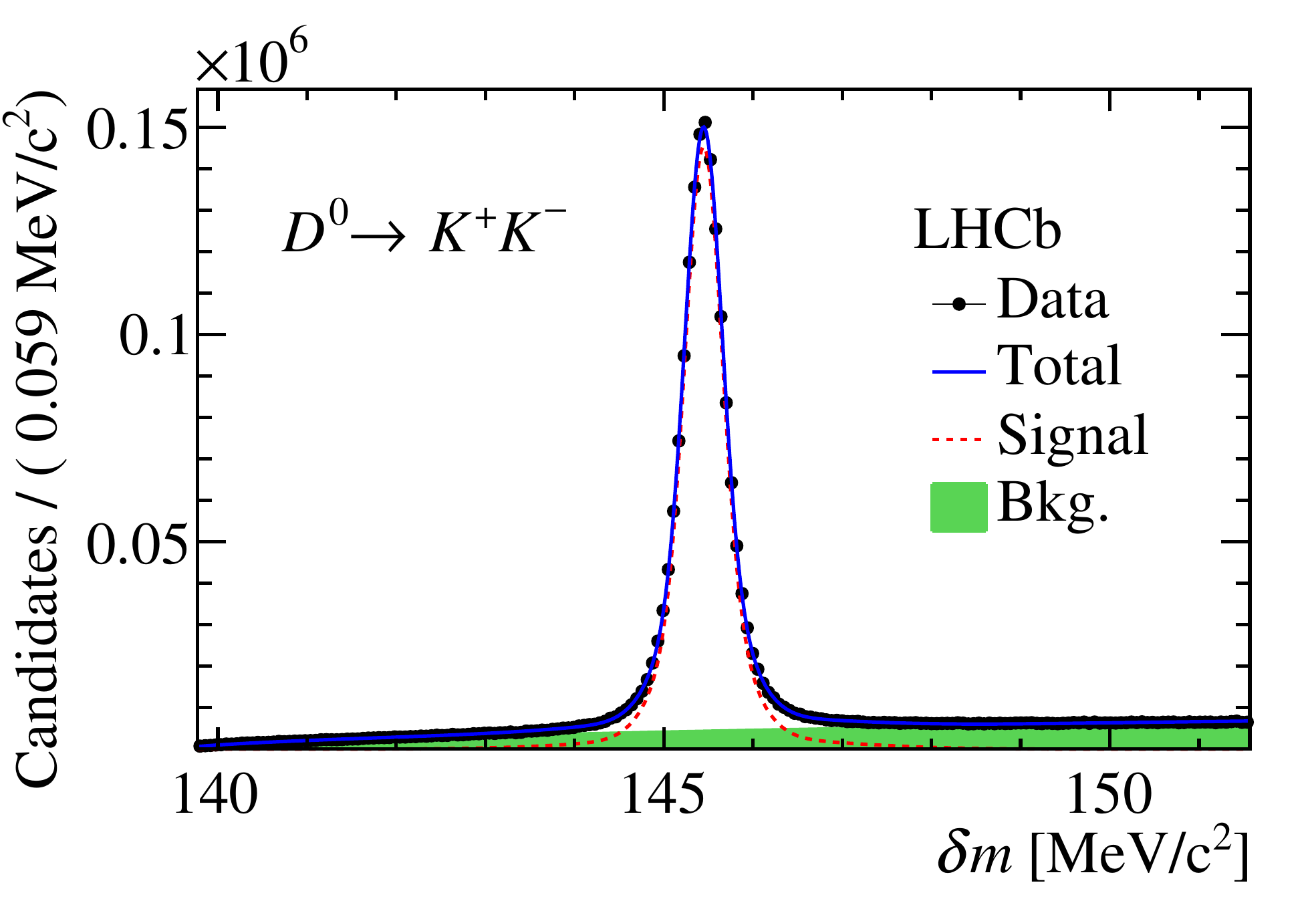}
\end{subfigure}
\begin{subfigure}{0.49\textwidth}
\includegraphics[scale=0.35,  trim = 0mm 0mm 0mm 0mm, clip]{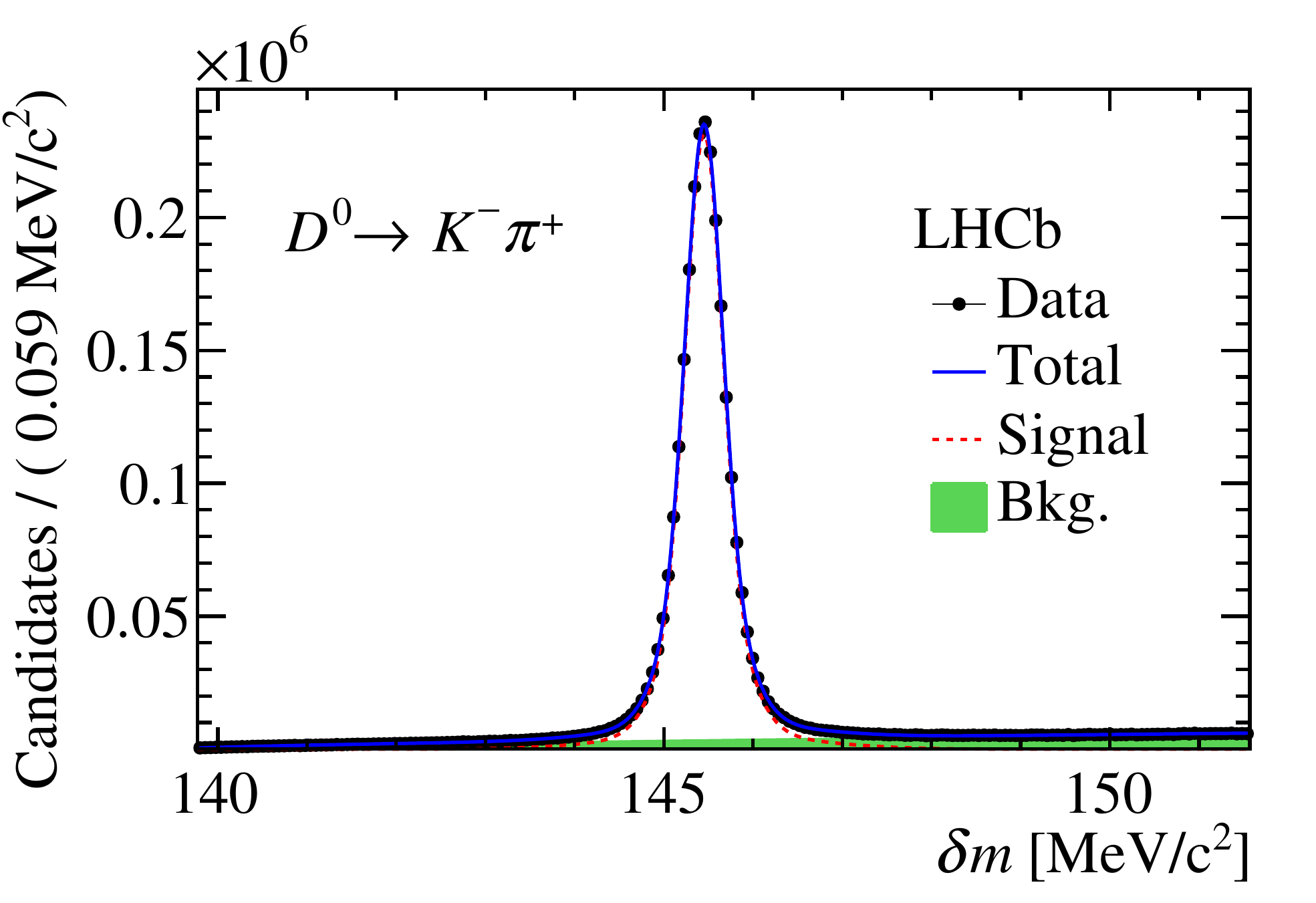}
\end{subfigure}
\begin{subfigure}{0.49\textwidth}
\includegraphics[scale=0.35,  trim = 0mm 0mm 0mm 0mm, clip]{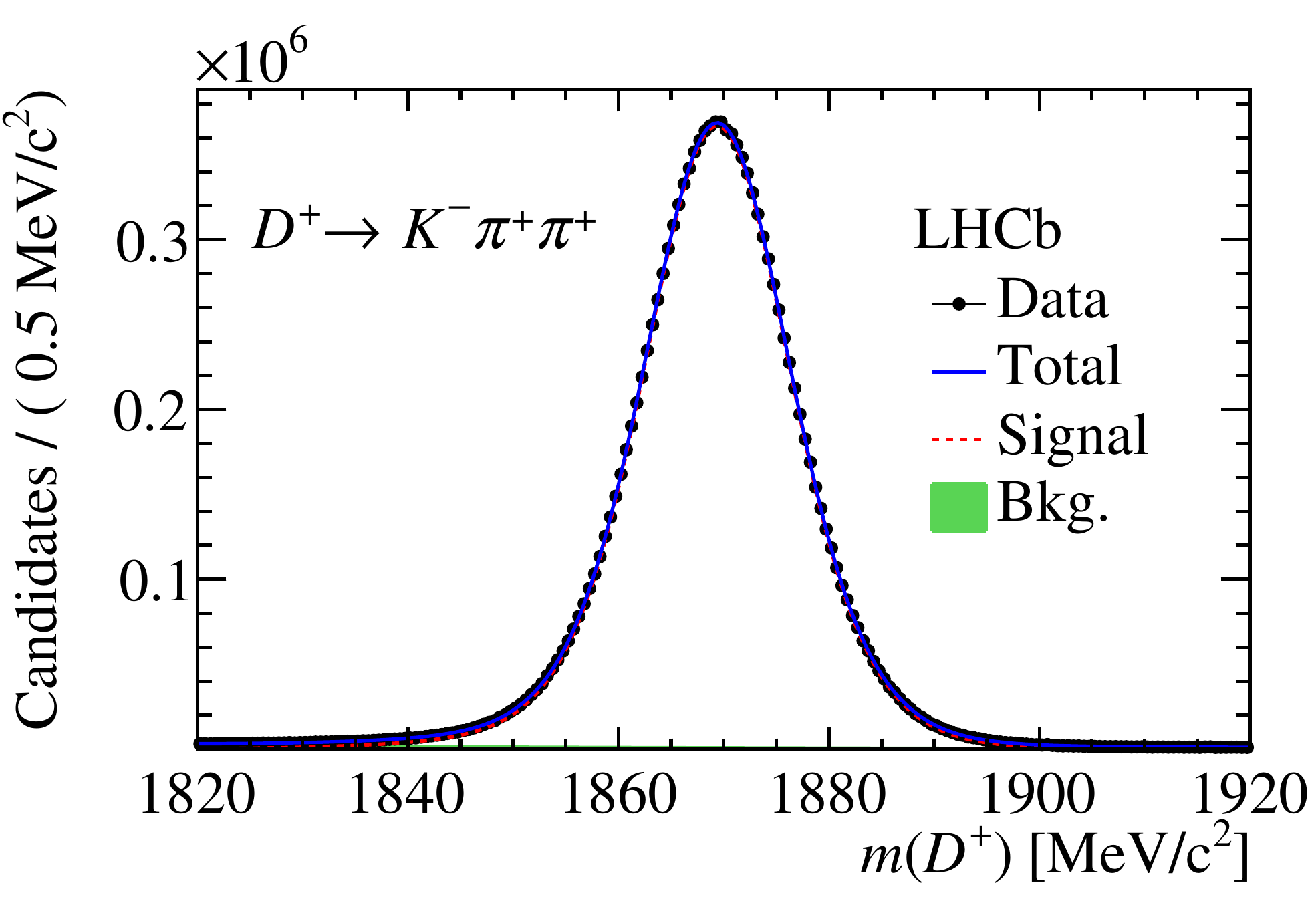}
\end{subfigure}
\begin{subfigure}{0.49\textwidth}
\includegraphics[scale=0.35,  trim = 0mm 0mm 0mm 0mm, clip]{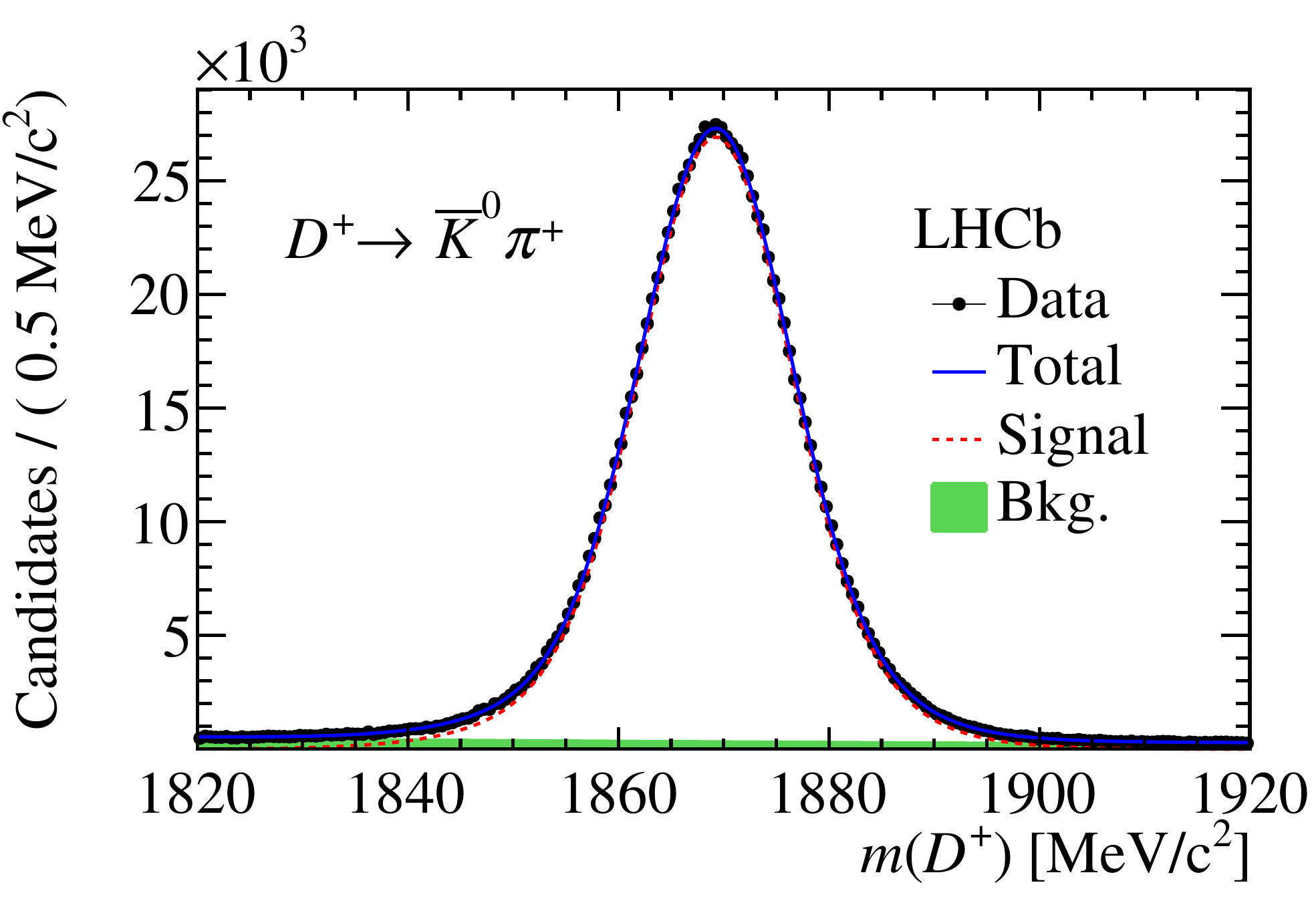}
\end{subfigure}
\caption{Fits to the $\deltam$ and to the $m(\Dp)$ distributions corresponding to the whole data sample and both flavours. Data samples after the kinematic weighting described in the text are used.}
\label{fig:fits}
\end{figure}

The raw asymmetries and the signal yields are determined from binned likelihood fits to the $\delta m = m(\Dstarp)-m(\Dz)$, distributions in the \Dz decay modes, and to the invariant mass distributions $m(\Dp)$ in the \Dp channels (see Fig.~\ref{fig:fits}). The yields of the decay modes, before and after the weighting procedure, are listed in Table~\ref{tab:yields}. 
\begin{table}[pt]
\caption{Signal yields of the four channels before and after the kinematic weighting. In the case of the weighted samples, effective yields are given.}
\centering
\begin{tabular}{l| c c c c|c}
 \textbf{Channel} &\textbf{Before weighting}& \textbf{After weighting}  \\
\hline
$\KK$  & $5.56\,\text{M}$ & $ ~~1.63\,\text{M}$ \\
$\KPi$ & $32.4\,\text{M}$ & $ ~~2.61\,\text{M}$ \\
$\KPiPi$   & $37.5\,\text{M}$ &$13.67\,\text{M}$ \\
$\KzbPi$ & $1.06\,\text{M}$ & $ ~~1.06\,\text{M}$ \\
\end{tabular}
\label{tab:yields}
\end{table}

The non-adequateness of the fit model, and the presence of peaking background can cause biases in the determination of individual raw asymmetries, and their effects are studied using pseudoexperiments, and simulated events. Possible presence of large local asymmetries can spoil the \acpkk measurement since Eq.~\ref{eq:AsymsKK} is valid only for small asymmetries. This systematic uncertainty is accounted for by excluding additional regions of the detector acceptance, and comparing the result to the nominal one.

The non-cancellation of detection and production asymmetries is accounted for by modifying the configuration of the procedure. Another way to asses this systematic contribution is by modelling the kinematic dependancies of the raw asymmetries, and combining them with per-event weights from data.
Presence of secondary charm decays can bias the measurement due to the different asymmetries related to the production of secondary \Dz decays. The uncertainty was calculated by measuring the fraction of secondary decays, and using the measured production asymmetries at LHCb as an input.
A systematic uncertainty on the calculation of the neutral kaon detection asymmetry is also taken into account.
All other investigated sources are found to be negligible.

A summary of the systematic uncertainties considered can be found in Table~\ref{tab:systs} and added in quadrature to obtain the overall systematic uncertainty.
\begin{table}[ph]
\caption{Systematic uncertainties from the different categories.
The quadratic sum is used to compute the total systematic uncertainty.
}
\label{tab:systs}
\centering
\begin{tabular}{l c}
\textbf{Category} & \textbf{Systematic uncertainty}\boldmath [$\%$]\\
\hline
Determination of raw asymmetries:\\

\hspace{1em}Fit model & $0.025$ \\

\hspace{1em}Peaking background & $0.015$ \\

Cancellation of nuisance asymmetries:\\
\hspace{1em}Additional fiducial cuts & $0.040$ \\

\hspace{1em}Weighting configuration & $0.062$ \\

\hspace{1em}Weighting simulation & $0.054$ \\

\hspace{1em}Secondary charm meson & $0.039$ \\

Neutral kaon asymmetry & $0.014$ \\
\hline 
\textbf{Total} & $0.10 $ 
\end{tabular}
\end{table}

The time-integrated \CP asymmetry in \KK decays measured using prompt charm decays is determined to be
\begin{align}
A_{\CP}^{\mathrm{prompt}}(\Km\Kp)=(0.14\pm0.15\stat\pm0.10\syst)\%.
\end{align}
This result can be combined with previous LHCb measurements of the same and related observables.
A combination with $A_{\CP}(\Km\Kp)$ measured using secondary charm decays~\cite{LHCb-PAPER-2014-013} yields

\begin{align}
A_{\CP}^{\rm comb}(\Km\Kp)=(0.04\pm0.12\stat\pm0.10\syst)\%.
\end{align}
Since the same \Dp decay channels are employed for the cancellation of detection asymmetries, the result is partially correlated with $A_{\CP}^{\mathrm{prompt}}(\Km\Kp)$. The statistical correlation coefficient is $\rho_{\rm stat}=0.36$. 
The difference in \CP asymmetries between \Dz\to\Km\Kp and \Dz\to\pim\pip decays, $\Delta A_{\CP}^{prompt}$, is measured at LHCb using prompt charm decays~\cite{LHCb-PAPER-2015-055}.
A combination of the measurement of $A_{\CP}^{\mathrm{prompt}}(\Km\Kp)$ with \DACP yields a value for 
\begin{align}
A_{\CP}^{\mathrm{prompt}}(\pip\pim)=A_{\CP}^{\mathrm{prompt}}(\Kp\Km)-\Delta A_{\CP}^{prompt}=(\AcpPiPi\pm\AcpPiPiStatErr\stat\pm\AcpPiPiSysErr\syst)\%.
\end{align}
The statistical correlation coefficient of the two measurements is $\rho_{\rm stat}=0.24$.

A combination with the measurement using secondary charm results in
\begin{align*}
A_{\CP}^{\rm comb}(\pim\pip)=(\AcpLhcbPiPi\pm\AcpLhcbPiPiStatErr\stat\pm\AcpLhcbPiPiSysErr\syst)\%.
\end{align*}

\begin{figure}[h]
\centering
\includegraphics[scale=0.5]{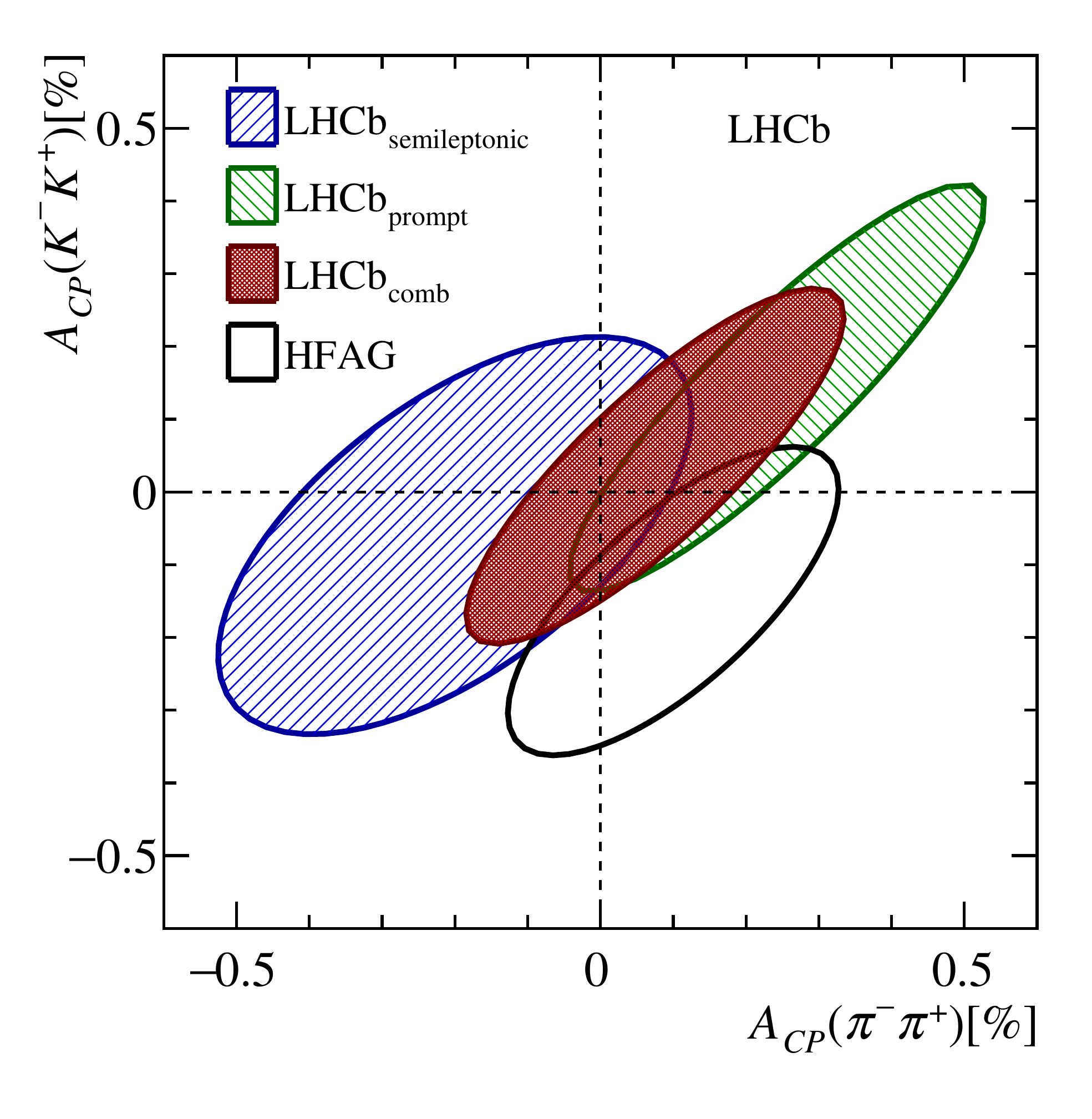}
\caption{Measurements of \CP violation asymmetries in \Dz\ra\Km\Kp and \Dz\ra\pim\pip decays.
Alongside the two LHCb measurements, presented in this Letter (green ellipse) and in Ref.\cite{LHCb-PAPER-2014-013} (blue ellipse), and their combination (red ellipse), the latest value of the Heavy Flavour Averaging Group \cite{HFAG} is shown (black ellipse). The latter already includes the measurement of \DACP with muon(pion)-tagged \Dz decays, using $3(1)\,$\invfb $pp$ collision data collected with the LHCb detector \cite{LHCb-PAPER-2014-013, LHCb-CONF-2013-003}. The $68\%$ confidence level contours are displayed where the statistical and systematic uncertainties are added in quadrature.}
\label{fig:AcpEllipses}
\end{figure}
Fig.~\ref{fig:AcpEllipses} shows the LHCb measurements of \CP asymmetry using both prompt and secondary $\Dz\to\Km\Kp$ and $\Dz\to\pim\pip$ decays. Additionally, the latest combined values of the Heavy Flavour Averaging Group \cite{HFAG} for these quantities are presented.
 
In conclusion, no evidence of \CP violation is found in the Cabibbo-suppressed decays \KK and \PiPi. These results are obtained assuming that there is no \CP violation in \Dz--\Dzb mixing and no direct \CP violation in the Cabibbo-favoured \mbox{\KPi}, \KPiPi and \KzbPi decay modes. The combined LHCb results are the most precise measurements of the individual time-integrated \CP asymmetries \acpkk and \acppipi from a single experiment to date.

\newpage
\section*{References}

\bibliography{main,LHCb-PAPER,LHCb-CONF,LHCb-DP,LHCb-TDR}

\end{document}